# Avalanche-like behavior of up-conversion luminescence by nonlinear coupling of pumping rates

M. V. KOROLKOV[1], I. A. KHODASEVICH[1], A. S. GRABTCHIKOV[1], D. MOGILEVTSEV[1*], E.V. KOLOBKOVA[2,3]

[1]*Institute of Physics NASB, Nezavisimosty Ave 68-2, 220072, Minsk, Belarus*
[2]*ITMO University, 49 Kronverskii Av., St. Petersburg, 197101, Russia*
[3]*Saint-Petersburg State Institute of Technology (Technical University), 26 Moskovskiy Av., St. Petersburg, 190013, Russia*

*Corresponding author: d.mogilevtsev@ifanbel.bas-net.by*

Here we report and discuss the avalanche-like up-conversion behavior in absence of the avalanche. We experimentally observed significant changes in the slope of the curve for the intensity dependence of up-conversion luminescence of erbium ions in green band (520–560 nm) on the pump intensity of the diode laser. Such changes are typical for the photon avalanche. However, the concentration of erbium ions is insufficient for an efficient exchange of energy between them, and the excitation of a photon avalanche is not possible. Using a simple three-level approximation of the up-conversion process model, we have shown that the observed avalanche-like luminescence process can also occur in the absence of a photon avalanche due to the non-linear relation between the efficiency of two pumping channels of erbium ion caused by intensity dependence of the absorption.

## 1. INTRODUCTION

Currently the frequency up-conversion processes are widely applied in biomedical and optical technologies. To name just a few applications, one can mention, for example, optical control of neuronal activity in optogenetics, fluorescence imaging, bioassays and labels techniques, the "theranostics" (a new term that reflects the simultaneous combination of both therapeutic and diagnostic capabilities) in medicine, up-conversion lasers, highly-efficient photovoltaic devices for heat sensors and collecting solar energy [1-6]. In the majority of these applications the up-conversion processes are realized upon the rare earth ions (REIs), which are the staple basis for modern efficient optical solid-state technologies, including fiber systems, optical communications, etc. [7-10].

Up-conversion processes with rare earth ions (REIs) can manifest so-called "photon avalanches" [11-14]. The photon avalanche process arises as a consequence of a nonlinear cross-relaxation energy transfer between ions. It is characterized by a drastic increase in the up-conversion signal intensity with increasing of the pump intensity. Here we report a curious experimental result: an avalanche-like increase in the up-conversion intensity in potassium gadolinium tungstate $KGd(WO_4)_2$ crystal (KGW) containing erbium ions ($Er^{3+}$) under the conditions when the usual avalanche is impossible. Also, we suggest an explanation of this phenomenon. Currently, the most frequently used method for describing the up-conversion processes in REI-doped material is the balanced equation method. This method allows one not only to qualitatively evaluate the data obtained, but also to make quantitative calculations. For that it is necessary to determine accurately the descriptive parameters of the material [11, 15-17]. Usually, for quantitative modeling one needs taking into account dozens of simultaneously occurring linear and nonlinear processes of population transfer between different ion energy levels and different ions, considering also absorption of pump radiation, and all the relaxation processes (both radiative and non-radiative). Thus, one needs for input several scores of descriptive parameters. Some of them can be measured in experiments, some are calculated, such as, for example, luminescent characteristics of REIs in crystals commonly estimated via Judd-Ofelt approach [18-20]. Such a complexity can be an obstacle to theoretical analysis aiming to optimize up-conversion processes by tuning the parameters. From other side, some basic characteristic features of up-conversion processes in REIs can be well captured even with a simple three-level model, which can indeed provide for a good qualitative description of the photon-avalanche effect [11].

With the help of a simple three-level model, we show that such a simple phenomenon as mutual nonlinear dependence of pumping rates of different REIs transitions, occurring in a natural way as a result of the absorption modification with increasing of the pump intensity, can give rise to quite non-trivial effects mimicking photon-avalanches. The reported experimental results corroborate this prediction.

As far as we know, in the theoretical studies of the up-conversion processes the linear relationship between the pumping rates of the ground and excited states have been commonly assumed. In this case, without of an avalanche, the dependences of the up-conversion power on the pump power also have a linear form in the double logarithmic representation [21]. At high powers, the slope of the curves may decrease due to the saturation effect. This dependence is completely different for the photon avalanche occurring in the up-conversion process [11-14]. For emergence of a photon avalanche, certain conditions need to be met, and the most important of them is a sufficient concentration of REIs for enabling energy exchange between ions. The second important condition for the formation of an avalanche is relatively low absorption efficiency of pump radiation in the ground state compared to the excited state.

The aim of this work is to demonstrate and to explain this avalanche-like behavior of the up-conversion intensity versus pump power in an experiment, when the first and the second condition might not be fulfilled, and therefore, photon avalanche is impossible.

## 2. EXPERIMENTAL RESULTS

Here we report an observation of seemingly counter-intuitive avalanche-like increase of up-conversion green (520-560 nm) luminescence with the raising of diode laser continuous-wave excitation at 970 nm for trace concentrations (< $10^{-4}$ wt.%) of erbium $Er^{3+}$ ions doped KGW. For such low concentrations the conventional avalanche can hardly be possible [7, 22].

We use the KGW crystal (PJSC "Inkrom" Russia) 4×4×2 mm cut along the crystallographic axis b ($N_p$), containing $Er^{3+}$ impurities. The Puma-970 diode laser (OJSC "Milon" Russia) is used as the source of continuous IR radiation. This source gives out a power of up to 10 W near 970 nm. Laser radiation was focused by an achromatic lens (*f* = 20 mm) into a beam with a diameter of about 200 μm, which provided a power density of up to 20 kW/cm² at the beam waist. Up-conversion radiation was decomposed on diffraction gratings of 1200 g/mm of an MS3540i spectrometer (Solar TII, Belarus) for detection of luminescence in the 510–570-nm range that gave spectral resolutions of 0.06 nm. The luminescence spectra were measured on a Spec 10:256 liquid-$N_2$-cooled CCD detector (Roper Scientific, USA). The radiation of up-conversion luminescence has been recorded in 90 ° geometry.

A typical $Er^{3+}$ up-conversion luminescence spectrum including two bands in the 520–560 nm region is shown in Fig. 1. The excitation of these bands by radiation with a wavelength in the region of 970 nm is a sequential process (see Fig. 2). The first step is absorption from the ground state corresponding to the transition $^4I_{15/2} \rightarrow {}^4I_{11/2}$, with the pumping rate $R_1$ in Fig. 2. The second step is absorption from the exited level $^4I_{11/2}$ to the level $^4F_{7/2}$ (with the pumping rate $R_2$), followed by relaxation and emission of green luminescence (520-560 nm) at the $^2H_{11/2} \rightarrow {}^4I_{15/2}$ and $^4S_{3/2} \rightarrow {}^4I_{15/2}$ transitions. Fig. 3 shows the Log-Log representation of the dependence of such up-conversion luminescence intensity on the pump power (experimental points are marked with "+"). Note a significant (almost two orders of magnitude) increase of luminescence intensity when the pumping power is changed from 1 to 8 W.

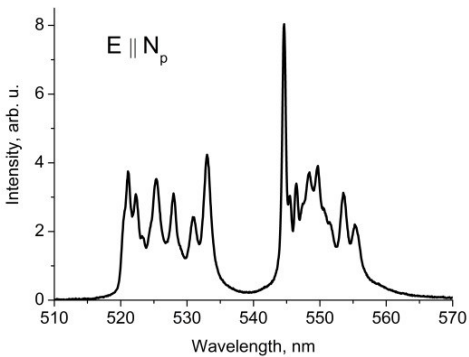

Fig. 1. Up-conversion luminescence spectrum for $Er^{3+}$:KGW excited by CW diode laser radiation at 970 nm.

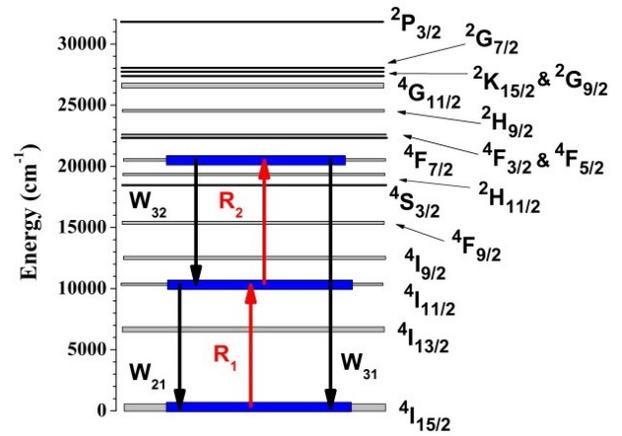

Fig. 2. Energy levels diagram for $Er^{3+}$:KGW and simplified three level scheme of the up-conversion process with 970 nm excitation.

Linear approximation of separate parts of the experimentally obtained data in Fig. 3 (+ points) shows that for different segments of the curve, in the Log-Log representation of this dependency, the parameter depicting the slope is n ≈ 1.5, 3.2 and 2.1, respectively. For the initial and end parts of the curve, n ≈ 2, which indicates the excitation of green emission by 2 pump photons [23], while in intermediate region parameter n is more than three, as it is usually observed, when a photon avalanche occurs [11, 14]. However this behavior cannot be caused by the excitation of a photon avalanche, since the concentration of erbium ions is insufficient for an efficient exchange of energy between them.

## 3. MODEL

To clarify the possible origin of this phenomenon, we have used a simple model of the up-conversion process. Fig. 2 shows a set of levels of $Er^{3+}$ ion in KGW relevant to the considered up-conversion luminescence process. Taking into account the stepwise excitation processes noted above ( $^4I_{15/2} \rightarrow {}^4I_{11/2}$, $^4I_{11/2} \rightarrow {}^4F_{7/2}$) incited by the radiation of a diode laser at 970 nm, we can impose an effective three-level scheme on this system of levels (see thick solid lines in Fig. 2). Three-level model presented in Fig. 2, is described by the following equations for populations, $n_i$, of levels involved in the model ($n_1 \to {}^4I_{15/2}$, $n_2 \to {}^4I_{11/2}$, and $n_3 \to {}^4F_{7/2}$):

$$\dot{n}_1 = \ldots + W_{21}n_2 + W_{31}n_3 - sn_1n_3$$
$$\dot{n}_2 = \ldots - (W_{21} + R_2)n_2 + W_{32}n_3 + 2sn_1n_3$$
$$\dot{n}_3 = \ldots - W_{31}n_3 - W_{32}n_3 - sn_1n_3$$

with parameters $W_{32}=23800$ c$^{-1}$, $W_{31}=2300$ c$^{-1}$ and $W_{21}=100$ c$^{-1}$ specific for Er$^{3+}$:KGW [22-25]. All numerical calculations presented in the article are performed with these parameter values. The parameter $s$ takes into account the efficiency of cross-relaxation energy transfer between ions [11, 14]. At low Er$^{3+}$ concentrations, as in our case, collective effects due to energy transfer between ions can be neglected ($s\to 0$). The dependences of the intensity of up-conversion luminescence proportional to the population of $n_3$ on the pumping rates $R_1$ and $R_2$ have been analyzed numerically and presented by surface $n_3=n_3(R_1,R_2)$ in Fig. 4. Cutting the surface $n_3(R_1, R_2)$ by the plane $R_2 / R_1 = $ const, gives the dependence of the luminescence intensity on the pump intensity, commonly used in the works of the subject (see, for example, green dashed line in Fig. 4). Typical dependencies in this case are shown in Fig. 5(a). Our analysis shows that under the condition $R_2 / R_1 = $ const, and typical parameters for Er$^{3+}$:KGW, the avalanche-like behavior of the dependence $n_3 = n_3(R_1; R_2 / R_1 = $ const) was observed only in the case of $R_2 \gg R_1$ and at $s > W_{31}$, in accordance with Ref. [11] (see lines 4 and 5 in Fig.5 (a)). In all other cases, a linear relationship was observed, which saturates with increasing pump power (lines 3, 2 and 1 in Fig. 5(a)).

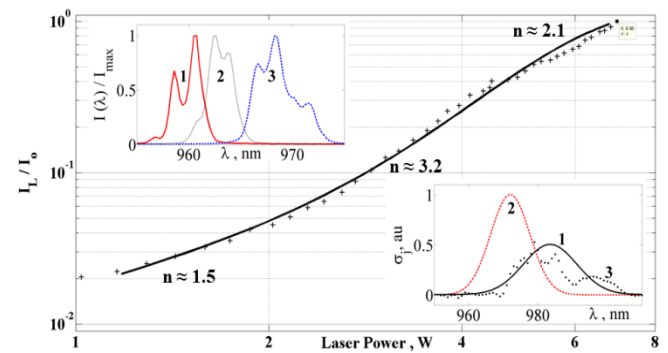

Fig. 3. The dependence of the intensity of up-conversion luminescence of $Er^{3+}$:KGW on the power of CW diode laser – experiment (+) and theory (line). The insets show the modification of the spectrum of the diode laser with increasing intensity (upper inset) and approximation by the Gaussian curves (lower inset) of the absorption cross section in the pump channels $R_1$ and $R_2$ (curves 1 and 2). The dots in the lower inset (curve 3) correspond to experimentally measured absorption for moderately high concentration (~1 wt. %) of $Er^{3+}$ ions in KGW crystal in the pump channel $R_1$.

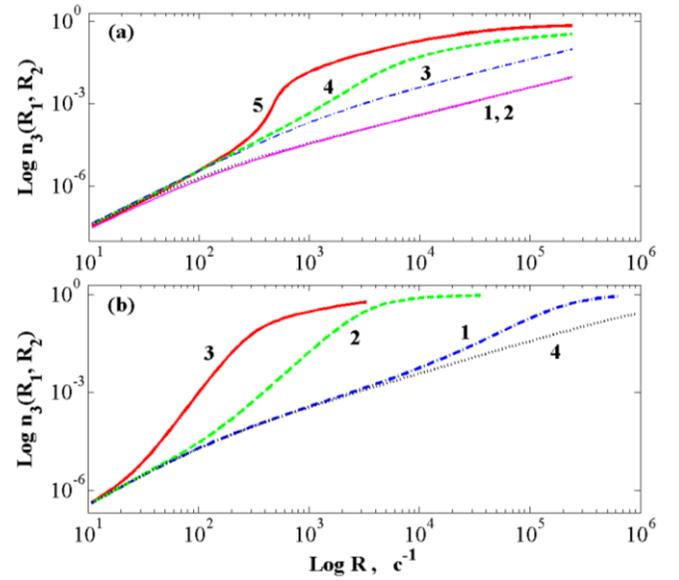

Fig. 5. (a) The dependence $n_3=n_3(R_1)$ for the case of $R_2 / R_1 = $ const. When the pumping rate in the first channel is significantly higher than in the second one ($R_1/R_2$=1000), the solid curve 1 corresponding to $s = 0$ and the dotted curve 2 corresponding to $s$=10000 almost coincide. In the opposite limit of $R_2/R_1$=1000 and sufficiently large $s > W_{31}$, an avalanche is observed for $s$=3000 (the dashed curve 4) and for $s$=10000 (the thick solid curve 5). For $s < W_{31}$ there is no avalanche (the dash-dotted curve 3 for $s$=1). (b) The dependence $n_3=n_3(R_1)$ for nonlinear relation between the rates $R_1$ and $R_2$ given by the equation $R_2=R_1(1+\alpha R_1^\beta)/\gamma$ for $\beta$=1 (the dash-dotted curve 1), $\beta$=2 (the dashed curve 2); $\beta$=3 (the thick solid curve 3); for $\alpha = 5\cdot 10^{-7}$. The dotted curve 4 describes the case without nonlinearity ($\alpha = 0$). For all the curves, the parameter $\gamma = 100$.

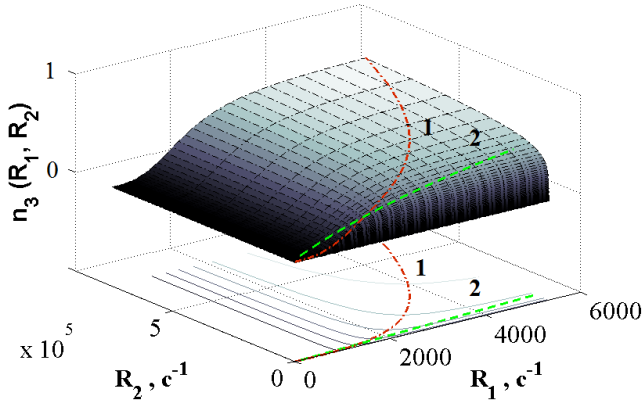

Fig. 4. Dependence of $n_3$ (which is proportional to up-conversion luminescence intensity) on the pumping rates $R_1$ and $R_2$ with linear and nonlinear coupling of pump rates (i.e., $R_1/R_2$=const. corresponds to the dashed curve 2, and $R_2/R_1=f(R_1)$ corresponds to the dash-dotted curve 1), and counter plot for them.

The case of nonlinear coupling of the pumping rates $R_1$ and $R_2$, shown by the dash-dotted curve in Fig. 4, can lead to a significant change in the shape of the curve as a result of more complex cutting of the surface $n_3=n_3(R_1,R_2)$. Typical shapes of the population $n_3 (R_1, R_2)$ in this case are shown in Fig 5 (b). Here, as an illustration, the results of numerical calculations of the effect of a polynomial nonlinear coupling of pump rates $R_2=R_1(1+\alpha R_1^\beta)/\gamma$ for three different orders ($\beta$=1,2,3) on the avalanche-like behavior of population $n_3$ in the stationary regime of excitation are presented.

First of all, for $\alpha = 0$, we expectedly have just a linear dependence $R_1 / R_2 = \gamma$ (the dotted curve 4 in Fig. 5(b)). Curiously, the different kinds of nonlinearity brings quantitatively different, but qualitatively rather similar avalanche-like behavior. One can see this behavior for quadratic, cubic and quartic nonlinearities (correspondingly, $\beta$=1, 2, 3). Notice that such a phenomenon occurs for $s = 0$ and $R_1$ possibly being of the same order as $R_2$.

It is remarkable that nonlinear coupling between the pumping rates naturally arises in experiments. This phenomenon can appear as a consequence of the spectral band shift of the pumping laser with an increase in the voltage applied to it (which was actually observed in our experiment). The upper inset in Fig. 2 shows the modification of the generation spectrum with an increase in output power from 0.66 W to 3.95 W and further to 9.2 W. In the first case, the average shift is about 4 nm, in the second it is about 6 nm. Such a change in the spectrum with an increase in the power of the diode laser leads to a change in both absorption from the ground (${}^4I_{15/2}$) and excited (${}^4I_{11/2}$) ionic states, which in turn changes the efficiency of both pump rates $R_1$ and $R_2$ proportional to the corresponding absorption cross sections. In this case, efficiency of the excitation processes depends on the values of the integral contributions averaged over the pump spectrum. Note that a modification of the diode laser emission spectrum is not the only possible effect leading to the appearance of a nonlinear connection between $R_1$ and $R_2$. The absorption spectra of the ground and excited states of REIs are shifted relative to each other [26,27] and may have their own structure (for more details, see, for example, Ref. [25-27]). So, for example, a shift of the pump frequency caused by temperature changes or nonlinear optical processes might also lead to a nonlinear connection between $R_1$ and $R_2$. The modification of the spectrum shown in the upper inset in Fig. 3 is essential for obtaining the avalanche-mimicking behavior in our experiment. We have tried to fit our experimental data varying the pump intensity and assuming that $R_1 \propto \sigma_1 P_{LD}$, $R_2 \propto \sigma_2 P_{LD}$, where $\sigma_1$ and $\sigma_2$ are the absorption cross sections of the lower and first excited levels. At room temperature, the typical features of variations of these absorption cross sections with increasing pump power of a diode laser $P_{LD}$ can be captured with Gaussian functions ( see, for example, Refs. [26,27]):

$$\sigma_j(P_{LD}) = b_j + c_j \exp\left[-\left(\frac{\Lambda(P_{LD})-\Lambda_j}{\Delta_j}\right)^2\right].$$

Here the shift of the center of the spectrum band of the laser diode (upper inset in Fig. 3) is assumed to be linearly dependent on the pump power: $\Lambda(P_{LD})$ - 959.2 nm ~ k·$P_{LD}$. Using these assumptions, we obtained the theoretical fit presented in Fig. 3 by solid line for the following set of parameters: $b_{1,2}$=0.002, $c_1$=0.04, $c_2$=0.024, $\Lambda_1$= 978 nm, $\Delta_1$= 6.3 nm, $\Lambda_2$= 969 nm, $\Delta_2$= 10.0 nm, and k=1.2. In Fig.3, the theoretical data are normalized with respect to the experimental data as to have the same maximum values.

Note that theoretical simulations similar to those used to produce results depicted in Figs. 4 and 5, were carried out also for a wide range of the model parameters (in particular, for more than 100% variation of parameters $W_{ij}$ of the system of equations describing our three-level model). The resulting surfaces $n_3 = n_3(R_1, R_2)$ did not have qualitative differences from the surface shown in Fig. 4. And the main result obtained above is, namely, the appearance of an avalanche-like behavior of the dependence $n_3 = n_3(R_1)$ due to the nonlinear dependence between the pumping rates $R_1$ and $R_2$, remains valid for them as well.

To clarify the nature of the dependence between the shifting spectra and the avalanche-like behavior, we have performed an additional experiment varying the pump power with variable absorbers (glass plates). So, the pump spectrum remained the same for different pump intensities. Expectedly, in this case the avalanche-like increase in the up-conversion intensity is not manifested. The modification of the pump spectrum leading to the modification in $R_1$ and $R_2$, was also confirmed by the measurement of the absorption spectrum in the ground state (curve 3 on the lower inset in Fig. 3). This figure shows that the absorption coefficient varies significantly in the region from 955 nm to 975 nm corresponding to the three average values of the spectra from the upper inset in Fig. 3 .

4. CONCLUSIONS

Concluding, we showed experimentally and explained theoretically how even for very low concentrations of erbium ions in a KGW crystal, the up-conversion process can exhibit the avalanche-like behavior in dependence of green luminescence intensity from intensity of the pumping diode laser at 960–970 nm. This behavior cannot be caused by excitation of a photon avalanche, because the concentration of erbium ions is insufficient for an efficient exchange of energy between them. Using a simple three-level model of this process, we demonstrate that an avalanche-like luminescence process can also occur due to nonlinear dependence of pump rates for different excitation channels of $Er^{3+}$ ion. The developed model demonstrates that this phenomenon has a general character and can be manifested in different schemes of up-conversion when the pump spectrum is intensity dependent. However, an interplay of this phenomenon with other effects, for example, with a true avalanche arising for a larger concentration of $Er^{3+}$ ions, is an open question.

**Funding.** Belarusian Republican Foundation for Fundamental Research (F15-116); Russian Scientific Foundation (19-13-00343 ); GPNI NAN Belarus "Convergence 2020".